\documentclass[showpacs,twocolumn,aps,floatfix,superscriptaddress]{revtex4}
\usepackage{amsmath,amsfonts,amssymb,eucal,graphicx,graphics,bm}

\begin{document}
\title{Exciting Hard Spheres}
\author{T.~Antal}
\affiliation{Program for Evolutionary Dynamics, Harvard University,
Cambridge, MA~ 02138, USA} 
\author{P. L. Krapivsky}
\author{S.~Redner}
\affiliation{Center for Polymer Studies \& Department of Physics, Boston
  University, Boston, MA~ 02215, USA}

\begin{abstract}
  We investigate the collision cascade that is generated by a single moving
  particle in a static and homogeneous hard-sphere gas.  We argue that the
  number of moving particles at time $t$ grows as $t^\xi$ and the number
  collisions up to time $t$ grows as $t^\eta$, with $\xi=2d/(d+2)$,
  $\eta=2(d+1)/(d+2)$, and $d$ the spatial dimension.  These growth laws are
  the same as those from a hydrodynamic theory for the shock wave emanating
  from an explosion.  Our predictions are verified by molecular dynamics
  simulations in $d=1$ and $2$.  For a particle incident on a static gas in a
  half-space, the resulting backsplatter ultimately contains almost all the
  initial energy.
\end{abstract}
\maketitle
\maketitle

The classical hard-sphere gas serves as a paradigm for a real molecular gas
for situations where quantum effects are negligible and where the
intermolecular potential is also negligible except when two molecules are in
close proximity.  This model has been successfully applied to understand many
statistical properties of equilibrium and non-equilibrium molecular fluids
\cite{fluid}.  

\begin{figure}[ht]
\centering
\includegraphics*[width=0.375\textwidth]{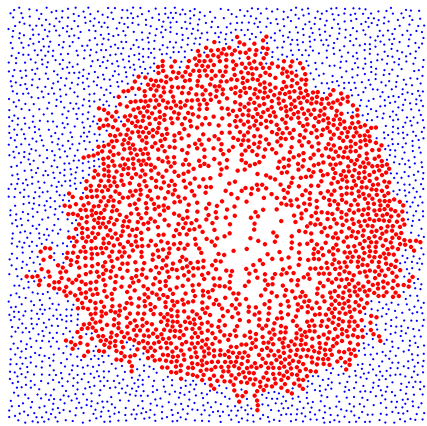}
\caption{(color online) Collision cascade in two dimensions at volume
  fraction $\rho=0.4$ and $t\approx 552$.  Moving spheres are shown actual
  size (red) and stationary ones at half radius (blue).  The cascade slowly
  moves rightward due to a particle initially at the center moving to the
  right.}
\label{dens}
\end{figure}

The hard-sphere gas consists of identical spherical particles that move along
straight-line constant-velocity trajectories that are interrupted by elastic
collisions with other spheres.  Despite its simplicity and considerable
research devoted to this model, many unexpected properties continue to be
discovered, such as the slow decay of velocity correlations \cite{AW} and the
anomalous statistics of collision events \cite{trizac}.  Here we investigate
a phenomenon of a somewhat different genre, namely, the response of a
stationary hard-sphere gas to the perturbation of a single moving particle.
After the first collision, two particles are moving and these, in turn,
collide with other particles, ultimately leading to a remarkably symmetric
cascade of moving particles (Fig~\ref{dens}).  In this collision cascade, we
argue that the number of moving particles at time $t$ grows as $t^\xi$ and
the total number collisions up to time $t$ grows as $t^\eta$, with
$\xi=2d/(d+2)$, $\eta=2(d+1)/(d+2)$, and $d$ is the spatial dimension.

The above setting is just the initial ``break shot'' in an infinite billiard
table.  In classical billiards theory, a wealth of beautiful phenomena has
been discovered about periodic and chaotic motions of a single particle that
collides elastically with a confining boundary \cite{T,G}.  In real
billiards, however, multiple particles move simultaneously.  While some
intriguing results exist about the collision dynamics of few-particle systems
\cite{few}, little is known quantitatively about infinite-particle systems,
such as the break shot that is the focus of this work.

The sudden injection of energy by the incident particle in the interior of a
zero-temperature gas can also be regarded as an explosion, and the collision
cascade is then analogous to the shock wave that propagates outward from the
initial detonation.  From hydrodynamic theory \cite{blast,landau}, basic
properties of the shock wave, such as its propagation velocity, and the
radial dependences of the density and temperature profiles behind the wave
are well understood.  Strikingly, the collision cascade in the hard-sphere
problem shares many quantitative properties of this shock wave.

Two natural characteristics of the cascade are $N_t$, the number of moving
particles at time $t$, and $C_t$, the total number of collisions up to time
$t$.  These quantities depend on the initial sphere locations, but numerical
simulations indicate that relative sample-to-sample fluctuations vanish in
the long time limit.  Therefore we focus on the number of moving particles
and number of collisions averaged over many initial conditions, $N(t)\equiv
\langle N_t\rangle$ and $C(t)\equiv \langle C_t\rangle$, respectively.

The number of moving particles can depend on time $t$, the density $n$, the
spatial dimension $d$, the initial velocity $v_0$ of the moving particle, and
the particle radii $a$: $N(t) = N[t, v_0, a, n, d]$.  Since the particle
number is dimensionless, dimensional analysis tells us that $N(t)$ should
depend on the three independent dimensionless parameters $T=n^{1/d}\,v_0t$,
$\rho = n \Omega_d a^d$ and $d$:
\begin{equation}
N = N[T, \rho, d],
\end{equation}
where $\Omega_d = \pi^{d/2}/\Gamma(1+d/2)$ is the volume of unit sphere in
$d$ dimensions.  The self-similar nature of the collision cascade then
suggests that $N(t)$ should have the power-law time dependence
\begin{equation}
\label{Nt-scal}
N \sim T^\xi~,
\end{equation}
with exponent $\xi$ that potentially depends on $\rho$ and $d$

We determine $\xi$ by writing a rate equation for the evolution of the number
of moving particles.  From Fig.~\ref{dens}, only particles within a shell of
thickness one mean-free path $\ell\sim v\tau\sim 1/(na^{d-1})$ from the
edge of the cascade at time $t$ can recruit additional moving particles in a
collision time $\tau$.  We now assume equipartition in which the initial
energy is shared approximately equally among the moving particles;
simulations clearly support this assertion.  At time $t$, each particle in
the cascade then has energy that is roughly $N^{-1}$ times the initial
energy, and root-mean-square velocity $v(t)\sim N^{-1/2}$.  Using the
observation from Fig.~\ref{dens} that the collision cascade is compact, the
number of particles within the collision shell is $n\,\ell\, R^{d-1}$.
Consequently, the rate equation is
\begin{equation}
\label{N-rate}
\frac{dN}{dt} \sim \frac{n\,\ell\, R^{d-1}}{\tau}
\sim n v \left(\frac{N}{n}\right)^{(d-1)/d}~.
\end{equation}
Now using $v\sim N^{-1/2}$ and solving the resulting differential equation
yields \eqref{Nt-scal}, with
\begin{equation}
\label{Nt}
\xi= \frac{2d}{d+2}~,
\end{equation}
a universal exponent that does not depend on $\rho$.  Notice also that $N$
itself does not depend on $\rho$.  We might anticipate that $\rho$ determines
the time at which asymptotic behavior sets in, with this crossover time
becoming long for small $\rho$.

As mentioned at the outset, the properties of the collision cascade can also
be determined by a hydrodynamically-based dimensional analysis.  The
perturbation caused by a single moving particle can be viewed as the
instantaneous release of energy $E=\frac{1}{2}mv_0^2$.  This point energy
source is akin to detonating an explosion, and the latter problem has an
elegant scaling solution \cite{blast,landau} as long as the pressure behind
the shock wave greatly exceeds that in front.  This condition is always
satisfied in our case, as the exterior gas has zero pressure.  In this
infinite Mach number limit, the radius $R(t)$ of the shock wave can only
depend \cite{blast} on the energy release $E$, the mass density $nm$, where
$m$ is the particle mass, and the time $t$.  Therefore the radius must scale
as
\begin{equation}
\label{Rt}
R(t)\sim (Et^2/nm)^{1/(d+2)}~,
\end{equation}
as this is the only variable combination with units of length.  The only
feature not determined by this dimensional analysis is the prefactor in
Eq.~\eqref{Rt} which requires solving the hydrodynamic equations of motion
\cite{blast,landau}.  Using the result for $R(t)$ in conjunction with
$E=\frac{1}{2}mv_0^2$ and $N=n\Omega_dR^d$, we again recover \eqref{Nt-scal}
\& \eqref{Nt}.

To obtain the number of collisions $C(t)$ up to time $t$, we note that in a
collision between a moving and a stationary particle one additional particle
moves \cite{head}, while in a collision between moving particles their number
remains the same.  Thus we have the trivial bound $C(t) \geq N(t)$, which is
valid for all times if we count the impulse that causes the initial particle
motion as a collision.  We again anticipate that the total number of
collisions scales algebraically with time $C \propto T^\eta$.

To determine $\eta$, we write the rate equation for $C$
\begin{equation}
\frac{dC}{dt}\sim \frac{N}{\tau}~
\end{equation}
that merely states that $N$ additional collisions occur per collision time.
We now use the fact that $\tau\sim \ell/v$ is proportional to $N^{1/2}/\rho$.
Together with Eq.~\eqref{Nt-scal} for $N$ and $T=n^{1/d} v_0 t$, we integrate
the rate equation to find
\begin{equation}
C \sim \rho^{(d-1)/d}\,T^{1+\xi/2}~.
\end{equation}
Therefore $\eta=1+\xi/2 = 2(d+1)/(d+2)$.

Although the initial trigger of the cascade is directional, the bulk motion
of the cascade is negligible for $d>2$.  Since the total momentum in the
cascade equals the initial momentum $mv_0=1$, when the cascade encompasses
$N$ particles, its center-of-mass velocity is given by $Nv_{\rm cm}=1$, or
$v_{\rm cm}\sim N^{-1}\sim T^{-\xi}$.  Thus the position of the
center-of-mass of the cascade is:
\begin{equation}
x_{\rm cm}(t) =\int_0^t v_{\rm cm}(t')\, dt' \sim 
\begin{cases}
t^{1/3} & d=1;\\
\ln t & d=2;\\
\mathrm{const.} & d>2.
\end{cases}
\end{equation}

\begin{figure}[ht]
\centering
\includegraphics[scale=0.425]{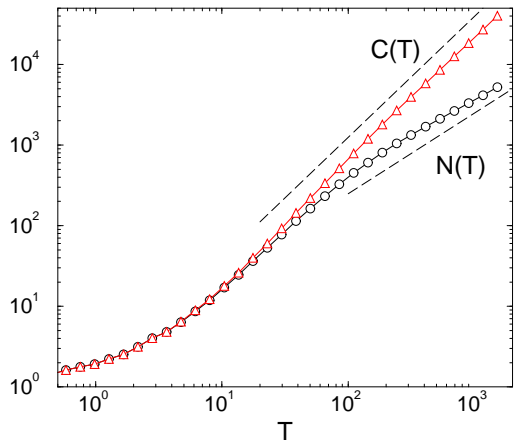}
\caption{Number of moving spheres $N(T)$ and number of collisions $C(T)$
  versus scaled time $T= n^{1/2}v_0t$ in two dimensions for volume fraction
  0.2 The dashed lines have slopes 1 and $3/2$ as predicted by theory.}
\label{CN-2d}
\end{figure}

To test our predictions, we perform molecular dynamics simulations on a gas
of perfectly elastic hard spheres.  In two dimensions, the system is
initialized by placing, one-by-one, a given number of spheres of the same
mass uniformly at random within a square box until a volume fraction $\rho$
is reached.  If overlap is created when a sphere is introduced, one of the
overlapping spheres is repositioned to eliminate this overlap.  Without loss
of generality we set the radius and the mass of spheres, and also the initial
speed of the initially-moving sphere that is in the middle of the box to 1.
The remaining spheres are initially at rest.  We stopped the simulations
before any moving sphere reaches the edge of the box.  Our data for the time
dependence of the average number of moving spheres and the number of
collisions is in good agreement with theory in the long-time limit
(Fig.~\ref{CN-2d}).  We also find that the exponents $\xi$ and $\eta$ are
close to the theoretical values of 1 and 3/2 respectively, and are
independent of the volume fraction $\rho$ for $0.1<\rho<0.45$.  It is worth
emphasizing that moving particles have their directions quickly randomized
for $d\geq 2$ so that the cascade becomes symmetrical about its center of
mass in spite of the directional initial perturbation.

The analogous problem in $d=1$ is pathological because there is no
``mixing''---two elastically colliding particles merely switch their
velocities.  By this construction, the collision cascade reduces to the
initial particle freely traversing the particle array.  To avoid this
pathology, we generalize slightly and allow for distributed particle masses.
This generalization leads to collisional velocity mixing and thus represents
a proper one-dimensional counterpart of our original system.  We specifically
consider uniformly distributed masses between 1 and $1-w$, with $w>0.2$ to
avoid the slow approach to asymptotic behavior if the mass distribution is
narrow.  We initially place particles at integer points in the domain $x\in
[-L,L]$; the particle at the origin is assigned a unit mass and velocity
$v_0=1$.

\begin{figure}[t]
\centering
\includegraphics[scale=0.425]{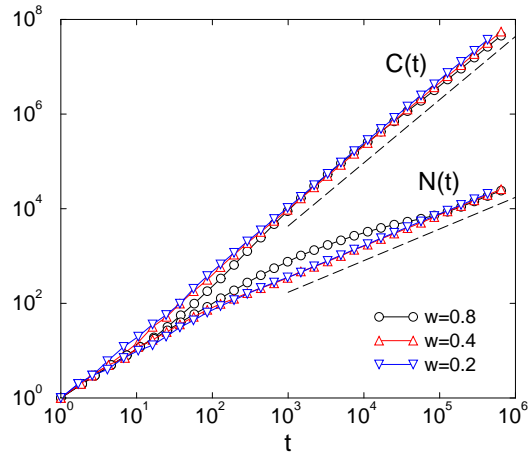}
\caption{(color online) Dependence of $N(T)$ and $C(T)$ versus scaled time
  $T= n v_0t$ in one dimension on a double logarithmic scale for
  different widths of the distribution of particle masses.  The asymptotic
  behavior is approached more slowly for narrower mass distributions.  The
  dashed lines have slopes 4/3 and 2/3 as predicted by theory. }
\label{CN-1d}
\end{figure}

With this initial condition, only particles in the region $x>0$ are excited
at early times.  However, the cascade gradually becomes more symmetrical!
After an initial transient, the ratio of the positions of the rightmost and
leftmost moving particles $|x_+/x_-|$ systematically decreases and approaches
a value that is close to 1 as $t\to\infty$.  This same approach to symmetry
occurs in the density profile of the gas.  The collision cascade also
satisfies the Rankine-Hugoniot conditions \cite{landau} that relate
discontinuities in thermodynamic properties of the gas across the shock front
in the infinite Mach number limit.  For example, the gas just behind the
leading edge of the cascade is twice as dense as that of the unperturbed gas
(for a monoatomic gas in one dimension; the density ratio is $d+1$ in $d$
dimensions), in agreement with Rankine-Hugoniot.  Similar results have been
found in large-scale numerical simulations of a planar shock in a hard-sphere
fluid \cite{shock}.  Finally, as shown in Fig.~\ref{CN-1d}, our simulation
results for $N(t)$ and $C(t)$ again agree with the theoretical predictions of
$\xi=2/3$ and $\eta=4/3$ in $d=1$.

\begin{figure}[ht]
\centering
\includegraphics[scale=0.7]{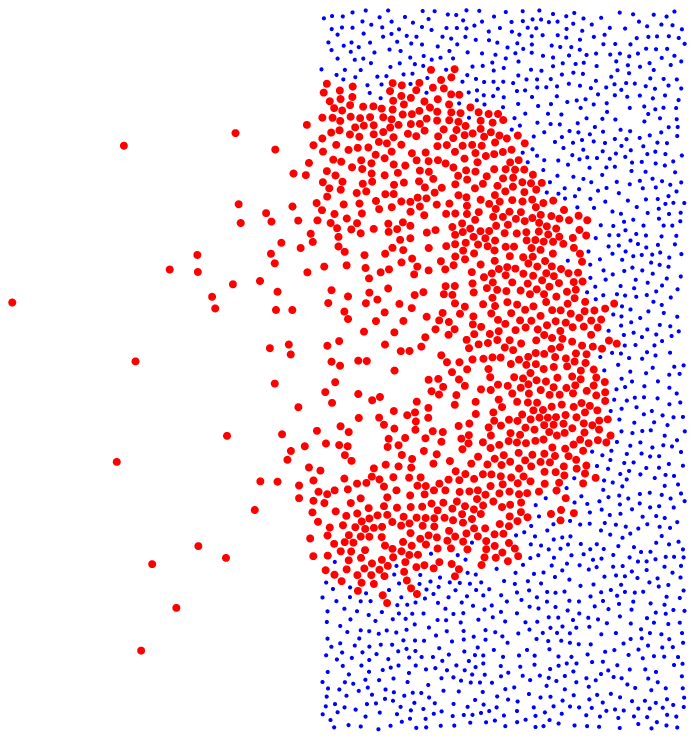}
\caption{(color online) Collision cascade in a two-dimensional half space at volume
  fraction $\rho=0.4$ and $t=400$ when a particle impacts from the left at
  $t=0$.}
\label{splash}
\end{figure}

We now study the ``splatter'' when a particle is normally incident on a
stationary gas in the half space $x\geq 0$ (Fig.~\ref{splash}).  Again a
collision cascade penetrates the gas in a manner similar to that of the bulk
system, but some particles are ultimately ejected backward.  Even though this
backsplatter is visually a small fraction of the particles in the cascade,
the energy contained in this backsplatter approaches value that is close to
(and it perhaps equal to) the initial energy.  In one dimension the situation
is more clear but still surprising.  At short times the incident particle
continues moving rightward as it recollides with the particle array.
Eventually, however, the incident particle is reflected so that it never
experiences another collision.  Particles $2, 3,\ldots$ undergo subsequent
and similar reflections so that a finite fraction of the particles are
ejected backward as a non-interacting ``fan''.  The kinetic energy in the
region $x>0$ slowly decreases with time and seems to vanish as $t^{-\beta}$
with $\beta \approx 0.1$, an observation that can be justified heuristically
\cite{future}.  Thus all the energy ultimately resides in the region $x<0$
and, in fact, is concentrated in the trailing edge of the backsplatter.  (A
related issue of optimal energy transmission through a finite array was
considered in \cite{trans}.)~  An important consequence of the vanishing
of the energy in the region $x>0$, is that Eq.~\eqref{Rt} now predicts that
the radius of the collision cascade grows as $t^{(2-\beta)/3}$ instead of
$t^{2/3}$ that arises from strict energy equipartition.
  
To summarize, the initial break shot on an infinite billiard table with a
finite density of stationary spheres creates an almost symmetrical collision
cascade within which excited particles share energy nearly equally.
Dimensional arguments determine the extent of the cascade as a function of
time.  The properties of the cascade are closely analogous to those for a
continuum shock wave that emanates from an explosion in a gas.  It will be
worthwhile to understand the limits of applicability of the continuum
hydrodynamic analogy, especially for one dimension.  The collisional splash
problem reveals unexpected features, the most prominent being that apparently
all of the energy is asymptotically carried by the backsplatter.  A major
challenge would be to prove that the energy transmitted to the initially
occupied half-space does indeed decay to zero; if so, to then determine the
decay law and whether the splatter is described within a hydrodynamic
framework.

We are grateful for financial support from NIH grant R01GM078986 (TA) and
Jeffrey Epstein for support of the Program for Evolutionary Dynamics at
Harvard University, as well as NSF grants CHE0532969 (PLK) and DMR0535503
(SR).


\begin{thebibliography}{99}
\bibitem{head} In a head-on collision between a moving and an identical
  stationary particle, the number of moving particles would not change;
  similarly, if particles meet with equal an opposite momenta they both would
  stop and the number of moving particles would decrease.  However, such
  events occur with probability zero.
\end{thebibliography}
\end{document}